# Extended strange metal and superconductivity in partially filled moiré Hubbard bands

Yiyu Xia[1,2*], Yichi Zhang[2,3*], Zhongdong Han[2,3*], Hongyuan Li[3,4], Brandon Bauer[5], Kenji Watanabe[6], Takashi Taniguchi[6], Wenjin Zhao[5], Jie Shan[1-4**], and Kin Fai Mak[1-4**]

[1]School of Applied and Engineering Physics, Cornell University, Ithaca, NY, USA
[2]Max Planck Institute for the Structure and Dynamics of Matter, Hamburg, Germany
[3]Laboratory of Atomic and Solid State Physics, Cornell University, Ithaca, NY, USA
[4]Kavli Institute at Cornell for Nanoscale Science, Ithaca, NY, USA
[5]Department of Physics and Astronomy, University of Delaware, Newark, DE, USA
[6]National Institute for Materials Science, Tsukuba, Japan

*These authors contributed equally
**Email: jie.shan@mpsd.mpg.de; kin-fai.mak@mpsd.mpg.de

**Moiré semiconductors provide a physical realization of the Hubbard model[1-3]. Bandwidth-tuned Mott transitions[4,5] and complex phase diagrams mimicking those of high-temperature superconductors[5] have been demonstrated near a weak Mott insulating state at half-band-filling. The platform also provides an opportunity to explore the physics of a partially filled Hubbard band at doping densities far away from the Mott insulator, a regime that has remained largely unexplored[6,7]. Here we report the emergence of an extended strange metal (ESM), in both doping density and twist angle, in a partially filled Hubbard band of twisted bilayer $WSe_2$. The ESM is in the clean limit, persists down to about 50mK, and spans over multiple symmetry-breaking ground states; it also appears only at small twist angles, where electronic correlations are strong[5,8,9]. These observations distinguish the state from the reported strange metals associated with quantum criticality[10-16] and favor the viewpoint that it is an intrinsic state of matter. We also report the observation of a new superconducting dome being developed from the ESM but far away from both the Mott insulating state and the van Hove singularity of the material. Our work opens the door to exploring the new regime of partially filled Hubbard bands in the clean limit.**

## Main

The Hubbard model is a simple theoretical model attempting to capture the effects of electronic correlations in materials with flat bands[6,7]. It describes electrons hopping in a lattice with a nearest-neighbor amplitude ($t$) and repelling each other with an onsite energy ($U$). Although the model appears simple and is a simplified representation of most real materials, in general, solutions to the model exist only near half-band-filling in either the strong ($\frac{U}{t} \gg 1$) or weak ($\frac{U}{t} \ll 1$) coupling limit[6,7]. Physical realizations of the model[17,18] can therefore provide crucial insights in the intermediate ($\frac{U}{t} \sim 1$) coupling regime, where it is believed to be relevant to the physics of high-temperature superconductors[19], and at generic doping densities away from half-band-filling, where solutions generally do not exist[6,7].

Moiré semiconductors provide an approximate realization of the Hubbard model with tunable doping density and electronic bandwidth ( $W \propto t$ ) [1-3]. Mott insulators[20,21] and kinetic magnetism[22,23] have been demonstrated in angle-aligned heterobilayers, which realize the Hubbard model towards the strong coupling limit[1]. Twisted homobilayer $WSe_2$ ($tWSe_2$) further allows efficient control of $W$ to the intermediate coupling regime through the twist angle ($\theta$) [9,24-29], thereby enabling the realization of a bandwidth-tuned Mott transition[4,5] and, in its vicinity, superconductivity[5,8,30-45] and complex phase diagrams mimicking those of high-temperature superconductors[5]. Whereas these studies provide important insights near half-band-filling, the regime far away from half-band-filling has remained largely unexplored. In particular, new phases of matter and phenomena may emerge in a partially filled Hubbard band with intermediate coupling strength, where the electronic kinetic energy and the Hubbard interaction are both important[6,19].

In this work, we report the emergence of an extended strange metal (ESM) and a new superconducting dome in a partially filled Hubbard band away from half-band-filling in $tWSe_2$ with twist angles $\theta \approx 3.4° - 3.7°$. As shown by earlier theoretical studies[24-27], the topmost moiré valence band of $tWSe_2$, when non-topological (valid for a range of twist angles, see Ref. [46]), can be mapped to a single-band triangular lattice Hubbard model with a widely tunable coupling strength $\frac{U}{t} \propto \frac{1}{\theta}$. A perpendicular electric field ($E$) can further tune the effective spin-orbit splitting in the moiré band structure and the location of the van Hove singularity (vHS). The combined effects of $\theta$ and $E$ thus provide a powerful control of the moiré Hubbard Hamiltonian. We fabricated dual-gated Hall bar devices of $tWSe_2$ with varying twist angles for electrical transport studies (Fig. 1a). The two gates allow independent control of $E$ and the doping density ($n$) measured in the lattice filling factor ($\nu \equiv \frac{n}{n_M}$), where $n_M$ is the moiré density. We also patterned contact and split gates to enable four-terminal measurements of the longitudinal ($\rho_{xx}$) and Hall ($\rho_{xy}$) resistivities as a function of $\nu$, $E$, temperature ($T$) and perpendicular magnetic field ($B$). Typical twist angle inhomogeneity over a 10μm-long channel length is less than 0.1° in our devices (Ref. [26] and Methods). See Ref. [5,30] for details on device fabrication, twist-angle calibration and electrical transport measurements.

**New superconducting dome near $\nu = 1/2$**

Figure 1d shows $\rho_{xx}$ as a function of $\nu$ and $E$ for four different samples with $\theta \approx 3.4° - 4.0°$ at $T = 30$mK and $B = 0$ T. The phase diagram can be divided into the layer-polarized and layer-hybridized regions (Fig. 1e). A Mott insulator with diverging $\rho_{xx}$ is observed at $\nu = 1$ inside the layer-hybridized region. Extended metallic regions of enhanced $\rho_{xx}$ surrounding the Mott insulator can also be identified; they correspond to a doped Mott insulating state with a small Fermi surface (see Extended Data Fig. 1 for $\rho_{xy}$ maps). Consistent with earlier studies[5,8,30,31], a superconducting region (SC1) with zero $\rho_{xx}$ is observed right next to the Mott transition near $\nu = 1$ inside the layer-hybridized region, and the electric field can induce a direct superconductor-to-insulator transition. With increasing $\theta$, the Mott transition moves to higher electric fields, and SC1 closely follows; the small Fermi surface regions also shrink. Away from $\nu = 1$, Wigner-Mott insulators at $\nu = 1/4$ and 1/3, as well as resistivity signatures of the vHS near $\nu = 0.78$ (more discussions below) can also be identified near $E = 0$mV/nm.

Interestingly, when the base electron temperature is lowered compared to earlier studies[5,30] (from $\approx 100$mK to $\approx 30$mK), a new superconducting region (SC2) near $\nu = 0.55$ and $E = 0$mV/nm is observed over a narrow twist-angle range $\theta \approx 3.4° - 3.7°$. [Signatures of a related superconducting region have also been reported recently in twisted trilayer $WSe_2$ (Ref. [47]).] SC2 weakens/disappears in both smaller (e.g. $\theta \approx 3.4°$) and larger (e.g. $\theta \approx 3.9°$ and 4.0°) twist-angle samples (see Extended Data Fig. 2 for additional data); it is replaced by a metallic and an insulating state for $\theta \gtrsim 3.8°$ and $\theta \lesssim 3.4°$, respectively (more discussions on the insulating state below).

We now focus on the $\theta \approx 3.5°$ device at $E = 0$mV/nm. Figure 2a shows $\rho_{xx}$ as a function of $\nu$ and $T$ at $B = 0$T. Line cuts at selected filling factors are shown in Fig. 2d. Both SC1 and SC2 show relatively sharp transitions to zero resistivity below the superconducting transition temperature ($T_c$). Compared to SC1, SC2 has a lower optimal $T_c \approx 120$mK at $\nu \approx 0.55$. To verify the superconducting nature of SC2, we examine the differential longitudinal resistance $\frac{dV}{dI}$ (at optimal doping) as a function of the DC bias current $I$ and $T$ ($B$) in Fig. 2b (Fig. 2c). The critical current vanishes continuously with increasing $T$ or $B$. Signatures of an interference pattern are also observed in the magnetic field dependent critical current. In Extended Data Fig. 3, we further extract the Cooper pair onset temperature, the Berezinskii-Kosterlitz-Thouless transition temperature and the upper critical field. The last quantity gives an estimate of the superconducting coherence length $\xi \approx 69$nm about 14 times the moiré period $a_M \approx 5$nm and about 17% of the electron mean free path $l \approx 410$nm (estimated from the residual resistivity).

The data in Fig. 2a,d also show that both superconducting domes are centered at fillings away from the vHS (centered at $\nu \approx 0.78$). [The observed location of the vHS is in good agreement with continuum-model band-structure calculations[27,48] (Fig. 1b,c).] The resistivity exhibits a peak at the vHS for $T \gtrsim 150$mK (due to the large density of states and/or enhanced scattering) and, for $T \lesssim 150$mK, a dip that corresponds to a phase transition to a symmetry-breaking ground state (see the line cut at $\nu \approx 0.78$ in Fig. 2d). The transition is likely driven by the enhanced interaction effect at the vHS. However, the state is not a superconductor, as shown by the non-zero resistivity at low temperatures and by the magnetoresistivity data in Extended Data Fig. 4. Its exact nature remains unclear at this point; further studies are required. Nevertheless, the results show that the superconducting domes are not induced simply by the large density of states at the vHS and cannot be explained by a pure electron-phonon mechanism or a weak-coupling Kohn-Luttinger mechanism, which would predict maximum $T_c$ at the vHS[36,44,45]. Such mechanisms also contradict the observed proximity of SC1 to the Mott transition regardless of the twist angle. The results thus suggest strong electronic correlations as the origin of the observed superconducting domes.

**Extended strange metal**

To further illustrate this point, we examine the normal-state transport properties at temperatures above the symmetry-breaking transition temperature $T_c$. (Hereafter, $T_c$ denotes a general symmetry-breaking transition temperature, whether into a superconductor or into the ordered state at the vHS, unless otherwise specified.) Figure 2d shows a $T$-linear resistivity for $T_c \lesssim T \lesssim 2$K over a wide doping density range $0.5 \lesssim \nu \lesssim 1$ (at $B = 0$T); the temperature dependence becomes slightly sublinear for $T \gtrsim 2$K. When the symmetry breaking transitions are suppressed by a small magnetic field of 50mT, the $T$-linear resistivity persists down to $T \approx 50$mK over the ranges $0.5 \lesssim \nu \lesssim 0.7$ and $0.87 \lesssim \nu \lesssim 1$, but Fermi liquid transport (i.e. $\rho_{xx} \propto T^2$) emerges near the vHS ($0.7 \lesssim \nu \lesssim 0.87$) at low temperatures (see Extended Data Fig. 5 for additional data). The small deviation from $T$-linear resistivity below $T \approx 50$mK is caused by the decoupling of the electron temperature from the lattice temperature.

The resistivity curves in the normal state can be described by $\rho_{xx} = \rho_0 + A \times T^{\alpha(T)}$. Here $\rho_0$ is the residual resistivity characterizing the disorder-limited transport, $A$ is a nearly temperature-independent amplitude and $\alpha(T)$ is a temperature-dependent exponent. (Note that $A$ is proportional to the Kadowaki-Woods ratio[49] for $\alpha = 2$.) By taking logarithmic derivatives of the experimental data (Methods), we extract the temperature dependence of $\alpha(T) \approx \frac{d\ln(\rho_{xx}-\rho_0)}{d\ln(T)}$ in Fig. 2e (valid for slow variations of $\alpha$ with temperature). The result is also summarized in a contour plot of $\alpha(T)$ versus $\nu$ and $T$ in Fig. 3f. The exponent is close to $\alpha = 1$ for $T_c \lesssim T \lesssim 2\text{K}$ (and $\alpha \lesssim 1$ for $T \gtrsim 2\text{K}$) over the doping density range $0.5 \lesssim \nu \lesssim 1$ at $B = 0\text{T}$. In contrast, Fermi liquid transport with $\alpha \approx 2$ is observed for $\nu \gtrsim 1$. The behavior remains similar at $B = 50\text{mT}$ except that a Fermi liquid exponent $\alpha \approx 2$ emerges near the vHS ($0.7 \lesssim \nu \lesssim 0.87$) at low temperatures (Fig. 2e and Extended Data Fig. 5).

In Extended Data Fig. 6, we extract the doping density dependence of $\rho_0$ and the resistivity dynamic range $\Delta\rho_{xx} \equiv \rho_{xx}(T = 2\text{K}) - \rho_0$. Both quantities show a peak near the vHS. The residual resistivity $\rho_0$ is small compared to $\Delta\rho_{xx}$ (also see Fig. 2d) and compared to the resistance quantum ($\frac{h}{e^2} \approx 25.8$ kΩ). (Here $h$ and $e$ denote Planck's constant and electron charge, respectively.) We also obtain a high electron mobility $\mu > 10^4 \text{cm}^2\text{V}^{-1}\text{s}^{-1}$ and a long mean free path $l \approx 400$ nm over the doping density range $0.5 \lesssim \nu \lesssim 1$ (Methods). The observed $T$-linear resistivity down to $T \approx 50\text{mK}$, the small $\rho_0$ and the high mobility over an extended doping density range (nearly a quarter of a full bandwidth) demonstrate the emergence of a clean ESM. The ESM here is distinct from earlier reports of strange metals in $tWSe_2$: It is much more extended in doping density than the strange metals reported in Ref. [4,5,50]; the $T$-linear resistivity also extends down to much lower temperatures than Ref. [4,50].

To further characterize the ESM, we examine the temperature-dependent magnetoresistivity and Hall resistivity near SC2 (Extended Data Fig. 7 and 8). The magnetoresistivity approximately follows a quadrature scaling, $\rho_{xx}(B,T) - \rho_{xx}(0,T) \propto \sqrt{(ak_BT)^2 + (b\mu_BB)^2}$, implying a scattering rate $\Gamma = \sqrt{(ak_BT)^2 + (b\mu_BB)^2}$. (Here $a$ and $b$ are dimensionless constants, and $k_B$ and $\mu_B$ are the Boltzmann constant and Bohr magneton, respectively.) Similar quadrature scaling has been reported in other strange metal materials[13]. Because of the small $\rho_0$ compared to $\Delta\rho_{xx}$, the magnetoresistivity also follows Kohler's rule reasonably well. Meanwhile, $\rho_{xy}$ shows a conventional weak temperature dependence, in contrast to the unusual $T^2$-dependent inverse Hall angle reported in some strange metals[5,13]. Note that due to the strong Zeeman splitting near the vHS[51] (Extended Data Fig. 4), analyses of the magnetoresistivity scaling collapse and the temperature dependence of $\rho_{xy}$ are limited to doping densities away from the vHS (see Methods for additional discussions).

**Twist-angle dependence**

Next, we examine the twist-angle evolution of the ESM and its correlation with SC2. Figure 3 shows $\rho_{xx}$ and the extracted exponent $\alpha$ as a function of $\nu$ and $T$ at $E = 0\text{mV/nm}$ and $B = 0\text{T}$ for four different twist angles (see Extended Data Fig. 9 for line cuts). Similar to the $\theta \approx 3.5°$ device, an ESM with $\alpha \lesssim 1$ is observed for $T \gtrsim T_c$ and $0.6 \lesssim \nu \lesssim 1$ in the $\theta \approx 3.4°$ device; the ESM is also in the clean limit (Extended Data Fig. 6). (A metal with $\alpha \lesssim 1$ remains qualified as a strange metal[52].) Different from the $\theta \approx 3.5°$ device, however, insulating states develop at $\nu = 1/2$ and $\nu = 1$ (we will examine the $\nu = 1/2$ state below); Fermi liquid behavior ($\alpha \approx 2$) emerges at low temperatures near the $\nu = 1/2$ insulating state and SC2 disappears (although SC1 remains); the crossover to a sublinear temperature dependence (i.e. $\alpha < 1$) also happens at a lower temperature.

In contrast, the $\theta \approx 3.65°$ device shows the same symmetry-breaking ground states as the $\theta \approx 3.5°$ device but the exponent away from the vHS starts to increase above $\alpha = 1$ at low temperatures (up to $\alpha \approx 1.3$), indicating a tendency to become a Fermi liquid with increasing $\theta$. Finally, Fermi liquid behavior appears away from the vHS at low temperatures for the $\theta \approx 3.9°$ device (although strange metal behavior, with exponent up to $\alpha \approx 1.1$, remains near the vHS). Correlated with the emergence of Fermi liquid transport, both superconducting domes disappear.

We construct the electronic phase diagram versus $\nu$ and $\theta$ at $E = 0$mV/nm in Fig. 4a,b (Methods). The electronic bandwidth $W$ increases monotonically with $\theta$. The presence of a Mott insulator at $\nu = 1$ in the $\theta \approx 3.4°$ but not the $\theta \approx 3.5°$ device shows that a Mott transition happens near the critical angle $\theta_C \approx 3.4° - 3.5°$, and SC1 appears near the Mott transition. The ESM emerges at $\theta \lesssim 3.8°$ and is likely driven by the strong electronic correlations in the intermediate coupling ($U \approx W$) regime near $\theta_C$. With sufficiently weakened coupling strength at large twist angles ($\theta \gtrsim 3.8° > \theta_C$), standard Fermi liquid behavior appears (although the strange metal behavior near the vHS persists). (The strange metal behavior near the vHS is originated from collisions between hot and cold electrons near the vHS[53].) Fermi liquid transport also appears above $\nu = 1$ at all twist angles.

Interestingly, SC2 also appears near $\theta_C$ and coexists with the ESM over the twist angle range $\theta \approx 3.5° - 3.7°$; it disappears when Fermi liquid transport emerges at larger twist angles. SC2 also disappears at $\theta \approx 3.4° \lesssim \theta_C$, where an insulating state dominates the phase diagram near $\nu = 1/2$. The nature of the $\nu = 1/2$ insulating state is examined in Fig. 4c-e and Extended Data Fig. 10. Magnetic hysteresis and a large anomalous Hall effect are observed at low temperatures. The temperature dependence of $\rho_{xx}$ and $\rho_{xy}$ (symmetrized and antisymmetrized at $B = \pm 10$ mT, respectively) shows a thermodynamic phase transition from a normal metal to a ferromagnetic insulator at a critical temperature $\approx 700$mK. The results thus suggest a competing relationship between the ferromagnetic insulator and SC2, and that SC2 emerges only when the ferromagnetic insulator is sufficiently weakened with increasing bandwidth or increasing electric fields in the $\theta \approx 3.4°$ device (Fig. 1d,e). These results, together with the fact that SC2 is away from the vHS, further support that the superconductor has its origin in strong electronic correlations rather than a pure electron-phonon pairing mechanism.

**Discussion and outlook**

We believe that the observed ESM in the clean limit is unusual and calls for a new understanding of the electrical transport in a partially filled Hubbard band near the Mott transition. First, the observed extended $T$-linear resistivity cannot be explained by electron-phonon scattering alone[54], which would predict such behavior only at temperatures above $\approx 1/3$ of the Bloch-Grüneisen temperature ($T_{BG}/3 \gtrsim 5$K and is weakly doping dependent); super-linear (i.e. $\alpha > 1$) temperature dependence is expected for $T < T_{BG}/3$. The observation of $T$-linear resistivity at temperatures (down to $\approx 50$mK) much smaller than $T_{BG}/3$ rules out this explanation. The appearance of Fermi liquid transport immediately above $\nu = 1$ is also inconsistent with this picture.

Second, it is unlikely that the effective Fermi temperature ($T_F$), below which $\alpha \approx 2$ appears, is less than 50-200mK over a wide doping density range[55]. As shown in Fig. 1b, continuum-model band-structure calculations show much higher single-particle Fermi temperatures $T_F \approx (60 - 110)$K for $\nu = (0.5 - 1)$. Although strong electronic correlations are known to suppress $T_F$[56], a $T_F < (50 - 200)$mK would imply the superconducting domes develop from a classical electron gas (i.e. $T_c > T_F$), which would violate a fundamental bound on $T_c$ for 2D superconductors[57].

Third, the observed ESM here is different from the reported strange metal associated with an extended quantum criticality[13,58,59], which often shows a positive correlation between $T_c$ and $A$ (Ref. [60]). Here, no obvious correlation between $T_c$ and $\Delta\rho_{xx}$ is seen (Extended Data Fig. 6); the ESM also spans over multiple symmetry-breaking ground states and is not tied to a particular extended quantum critical region. The ESM here is also different from that reported in twisted bilayer graphene[61-63] in several ways: 1) $\rho_0$ here is negligible compared to both $\frac{h}{e^2}$ and $\Delta\rho_{xx}$, but $\rho_0$ in twisted bilayer graphene is over an order of magnitude higher and is comparable to $\Delta\rho_{xx}$, implying the importance of disorder effects; 2) no quadrature scaling was found in twisted bilayer graphene; 3) the ESM here appears over a range of twist angles near the bandwidth-tuned Mott transition while that in twisted bilayer graphene appears only in the immediate vicinity of the magic angle.

Fourth, the ESM in general does not appear to be compatible with the scenario of non-Fermi liquids associated with a quantum critical point[10-12,14-16] (even a substantially broadened one in the strong disorder limit). Substantial disorder broadening can be ruled out by the observation of sharp quantum criticality at finite electric fields in the same device (Extended Data Fig. 11). As detailed above and in Methods, our devices are also in the low-disorder limit with weak disorder broadening ($\delta\nu \approx 0.02$ versus the extended range $\Delta\nu \approx 0.5$), small $\rho_0$ and high electron mobility.

Moreover, we want to point out that a theory for the ESM here should also be able to account for 1) the observed quadrature scaling, which suggests the absence of an intrinsic energy scale in the system; 2) the approximate obedience to Kohler's rule; 3) the weak temperature dependence of $\rho_{xy}$ as in conventional metals. Note that the last observation implies a single transport scattering time while the $T^2$-dependent inverse Hall angle observed in some strange metals requires the introduction of both the transport and Hall scattering times[64,65].

Interestingly, the only theoretical model we are aware of that appears to be compatible with our experiment is one that involves Umklapp scattering between the itinerant holes of a hole-doped Mott insulator and the spinon Fermi surface of a quantum spin liquid[52]; such quantum spin liquid could emerge near a bandwidth-tuned Mott transition in a triangular lattice[66], a situation close to our experimental setup[67]. Specifically, the model predicts non-Fermi liquid transport (including $\alpha < 1$) in the clean limit (i.e. $\rho_0 \to 0$) over an extended range of doping densities; it also predicts the obedience to Kohler's rule and a conventional Hall response, all of which are consistent with our experimental observations. (However, we emphasize that we do not attempt to claim experimental evidence of a quantum spin liquid.) Future thermopower measurements that could help extract the transport scattering rate will allow tests of a potential Planckian bound[68] in the ESM.

## Methods

### Determination of the power-law exponent $\alpha$.

To characterize the temperature dependence of $\rho_{xx} \approx \rho_0 + A \times T^{\alpha}$, we took a logarithmic derivative of the measured resistivity, $\alpha(T) \approx \frac{d\ln(\rho_{xx}-\rho_0)}{d\ln(T)}$, as described in the main text. First, we determined the residual resistivity $\rho_0$ at each filling by fitting the measured $\rho_{xx}(T)$ to a second-order polynomial, $\rho_{xx}(T) = \rho_0 + a_1 T + a_2 T^2$, over a low-temperature window. (Here $a_1 > 0$ and $a_2 > 0$ are fitting parameters.) The fitting window was chosen to lie above the transition temperature $T_c$ of the symmetry-breaking state (if it exists) while containing enough temperature points for a smooth and noise-robust extraction of $\rho_0$; a typical window was from $T_c$ to 0.8K.

Next, to obtain $\alpha(T) \approx \frac{d\ln(\rho_{xx}-\rho_0)}{d\ln(T)}$, we plotted $\ln(\rho_{xx} - \rho_0)$ against $\ln(T)$ and performed a linear fit to the data at each temperature over a narrow temperature window (typically containing about 10 temperature points); $\alpha(T)$ is the slope of the fit. This procedure substantially reduces the noise in the extracted $\alpha(T)$ compared to taking a direct numerical derivative $\frac{d\ln(\rho_{xx}-\rho_0)}{d\ln(T)}$ on the measured data (due to the noise in the measured $\rho_{xx}$).

Note that the extracted $\alpha(\nu, T)$ is rather insensitive to a small uncertainty ($\delta\rho_0$) in the extracted values of $\rho_0$ (except at the lowest temperature, $T \lesssim 100\text{mK}$) because $\rho_{xx} - \rho_0$ is much larger than $\delta\rho_0$ in most of the measured temperature range so that small variations in $\rho_0$ have a negligible effect on $\ln(\rho_{xx} - \rho_0)$. Only at the lowest temperatures, where $\rho_{xx} - \rho_0$ becomes comparable to $\delta\rho_0$, does the extraction become somewhat sensitive to $\delta\rho_0$. In general, the extraction of $\alpha(\nu, T)$ is reliable for $T \gtrsim 100\text{mK}$.

### Construction of the twist angle dependent phase diagram.

To construct the twist angle dependent phase diagram in Fig. 4b, we first combined the $\rho_{xx}$ data as a function of $\nu$ (at $E = 0\text{mV/nm}$, $B = 0\text{T}$ and $T = 30\text{mK}$) from the six devices measured in this study (Extended Data Fig. 2), and plotted a contour map of $\rho_{xx}$ versus $\nu$ and $\theta$ (Fig. 4a). We then read out the boundaries of the different states using this contour map and denoted the boundaries by symbols in the phase diagram (Fig. 4b). Guided by these symbols, we manually extrapolate the boundaries to obtain the region occupied by each state. The typical uncertainty in $\theta$ from such manual extrapolation is less than 0.1°.

The Fermi liquid and ESM regions were constructed based on the temperature dependence data in Fig. 3. Below $\nu = 1$, the 3.9° and 4.0° devices exhibit strange metal behavior only in the vicinity of the vHS and Fermi liquid behavior elsewhere, whereas devices with smaller twist angles (3.4°, 3.5° and 3.65°) show an ESM region even away from the vHS. Above $\nu = 1$, all devices exhibit Fermi liquid behavior. Note that unlike the sharp boundaries for the different symmetry-breaking states, the Fermi liquid and ESM regions are connected through smooth crossovers.

### Zeeman splitting of the vHS and its effect on magneto-transport.

Due to the large magnetic susceptibility near the vHS, a small magnetic field (< 10mT) is sufficient to induce a strong Zeeman splitting (Ref. [29,51]). (Note that the large magnetic susceptibility is also known to induce a ferromagnetic instability near the vHS in $tWSe_2$ samples[19] with small twist angles $\theta \lesssim 3°$.) As shown in Extended Data Fig. 4a, the vHS is quickly split into two branches under a magnetic field, with one moving towards higher fillings and the other towards lower fillings.

The strong Zeeman splitting has a significant effect on the magnetoresistivity near the vHS. Specifically, it gives a large negative magnetoresistivity for a range of fillings close to the vHS. Even for fillings away from the vHS, although a normal positive magnetoresistivity is first observed at low magnetic fields, a high enough field would ultimately lead to a large enough Zeeman splitting and induce a negative magnetoresistivity. Consequently, over a broad filling range (roughly $\nu \approx 0.55 - 1$, depending on the magnetic field strength), the magnetoresistivity is substantially influenced by the Zeeman splitting of the vHS and the associated modification of the band structure, rendering scaling analyses of the magnetoresistivity unreliable except at fillings far away from the vHS and at relatively weak magnetic fields (Extended Data Fig. 7).

The strong Zeeman splitting of the vHS also poses a challenge for the extraction of the weak-field Hall coefficient near the vHS because a truly weak magnetic field (that does not perturb the band structure) would yield unmeasurably small $\rho_{xy}$ near the vHS. For instance, the Zeeman splitting has already modified the band structure near the vHS substantially even at $B = 10$mT, greatly limiting the doping density range for a reliable measurement of $\rho_{xy}$; only at fillings far away from the vHS does a measurement of the weak-field $\rho_{xy}$ become possible (Extended Data Fig. 8). Similarly, the extraction of the quasiparticle effective mass from quantum oscillations around the vHS is difficult, as the magnetic fields required to resolve these oscillations (typically > 0.4T) are much larger than 10mT.

**Evaluation of disorder level**

To assess the disorder level in the ESM, we estimated the electron mobility $\mu$ and mean free path $l$ from the residual resistivity $\rho_0$ (small compared to $\Delta\rho_{xx}$) and carrier density $n$ in Extended Data Fig. 6. Away from the vHS, $\mu$ and $l$ can be estimated using the Drude formula $\mu = \frac{1}{ne\rho_0}$ and $l = \frac{\hbar k_F}{ne^2\rho_0}$, respectively (with $k_F = \sqrt{\pi n}$ and $k_F = \sqrt{2\pi(2n_M - n)}$ for $\nu < \nu_{\text{vHS}}$ and $\nu > \nu_{\text{vHS}}$, respectively). ($\nu_{\text{vHS}} \approx 0.78$ is the filling factor at the vHS.) Across different fillings and twist angles, we obtain $\mu$ on the order of $10^4$cm$^2$V$^{-1}$s$^{-1}$ and $l$ on the order of 100nm, demonstrating that the strange metal is in the clean limit.

The high sample quality is further supported by the observation of Shubnikov–de Haas oscillations at magnetic fields as low as 0.4 T (Extended Data Fig. 12) in the 3.65° device. From the onset field $B_{onset}$, we estimate a quantum mobility of $\mu_q \approx \frac{1}{B_{onset}} \approx 2.5 \times 10^4$ cm$^2$V$^{-1}$s$^{-1}$, consistent with the above estimated field-effect mobility.

Furthermore, our previous study[30] demonstrated the moiré density inhomogeneity in the 3.65° device to be around 2.5% across a $10\mu m$-long channel, corresponding to a twist-angle variation of approximately 0.05°. Altogether, our devices are in the low-disorder and homogeneous limit, supporting the observation of strange metal behavior in a clean system.

**Acknowledgement**
We thank Piers Coleman, Steven Kivelson, Gabriel Kotliar, Patrick Lee, Andrew Millis, Hoi Chun Po and Chandra Varma for helpful discussions.

## Figures

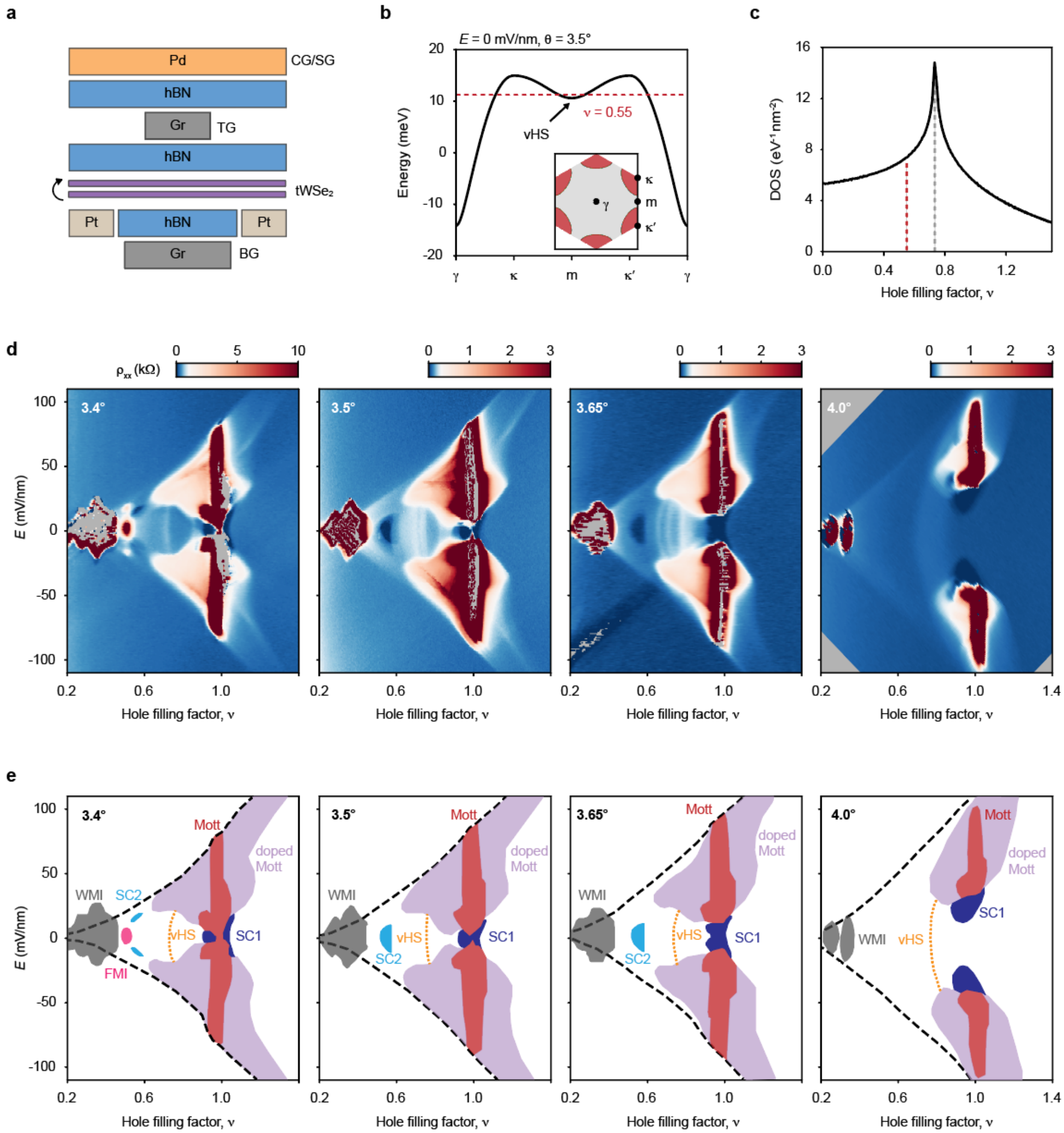

**Figure 1 | Dual-gated tWSe₂ devices with varying twist angles. a,** Schematic cross section of a dual-gated tWSe₂ device. CG, SG, TG and BG denote contact gate, split gate, top gate and bottom gate, respectively; Gr and Pd are few-layer graphite and palladium gate electrodes, respectively; Pt is the platinum contact electrode to tWSe₂; hBN stands for hexagonal boron nitride. **b,** Continuum-model band-structure of the topmost moiré valence band at $E = 0\text{mV/nm}$ and $\theta = 3.5°$. The arrow denotes the van Hove singularity (vHS); the dashed line marks the Fermi level at $\nu = 0.55$. Inset: Hole Fermi surface (red-filled) at $\nu = 0.55$. **c,** Corresponding density of states (DOS) as a function of $\nu$ at $E = 0\text{mV/nm}$. Dashed lines mark the positions for $\nu = 0.55$ (red) and the vHS (grey). **d,** Longitudinal resistivity $\rho_{xx}$ as a function of $\nu$ and $E$ at $T = 30\text{mK}$, $B = 0\text{T}$ and

$\theta = 3.4° - 4.0°$. **e,** Electronic phase diagrams constructed from the data in **d**. Grey: Wigner-Mott insulator (WMI); Pink: ferromagnetic insulator (FMI); Dark blue: superconductor near $\nu = 1$ (SC1); Light blue: superconductor near $\nu = 0.55$ (SC2); Red: Mott insulator (Mott); Purple: metallic state with a small Fermi surface (doped Mott); Yellow dotted line: location of vHS. Black dashed lines separate the layer-hybridized and layer-polarized regions.

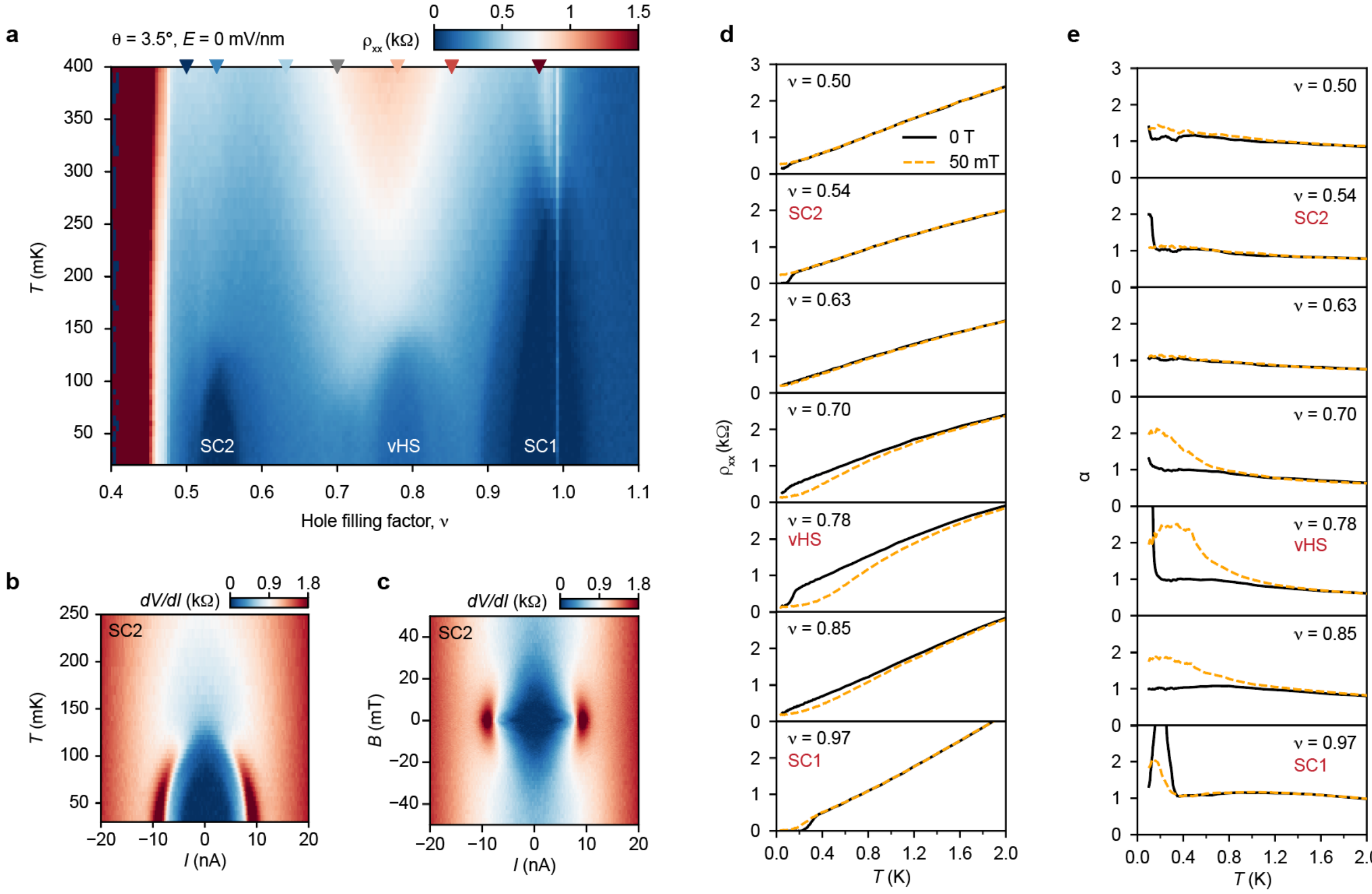

**Figure 2 | Extended strange metal and superconductivity in 3.5° tWSe₂. a,** $\rho_{xx}$ as a function of $\nu$ and $T$ at $B = 0$T. The locations for SC1, SC2 and vHS are labelled. **b,c,** Differential resistance $dV/dI$ at $\nu \approx 0.55$ as a function of DC bias current ($I$) and $T$ at $B = 0$T (**b**) and as a function of $I$ and $B$ at $T = 30$mK (**c**). **d,e,** $\rho_{xx}$ (**d**) and power-law exponent $\alpha$ (**e**) as a function of $T$ at selected fillings (marked by the triangles in **a**) under $B = 0$T (black solid line) and $B = 50$mT (yellow dashed line). $\alpha$ was extracted by the logarithmic derivative $\alpha(T) \approx \frac{d\ln(\rho_{xx}-\rho_0)}{d\ln(T)}$ (see Methods). All data were acquired at $E = 0$mV/nm.

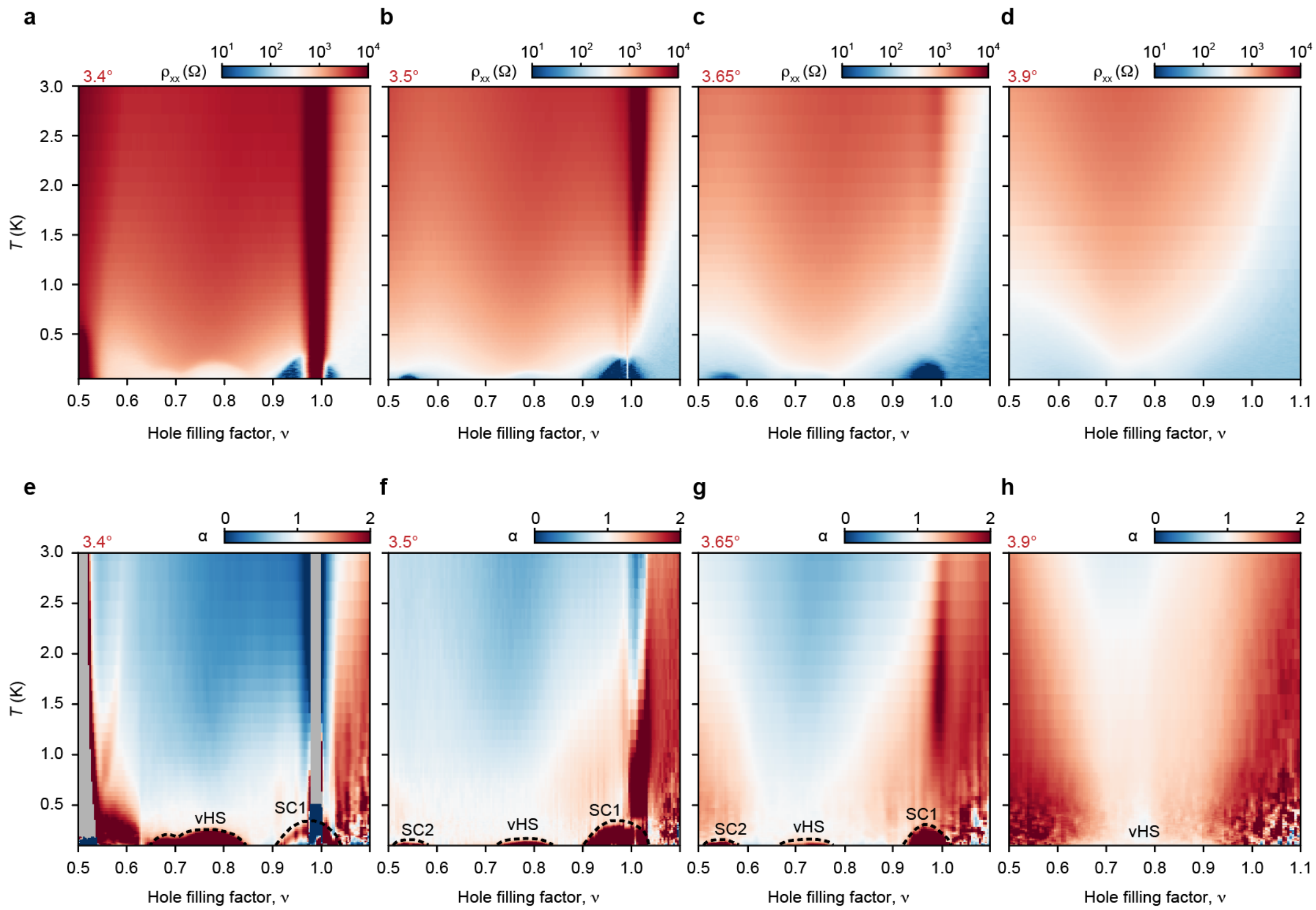

**Figure 3 | Twist angle evolution of the extended strange metal. a-d,** $\rho_{xx}$ as a function of $\nu$ and $T$ at $E = 0$mV/nm and $B = 0$T for $\theta = 3.4°$(**a**), 3.5°(**b**), 3.65°(**c**) and 3.9°(**d**). **e-h,** $\alpha$ as a function of $\nu$ and $T$ extracted from the data in **a-d**. The symmetry-breaking transition temperatures ($T_c$) are plotted for SC1, SC2 and vHS (dashed lines).

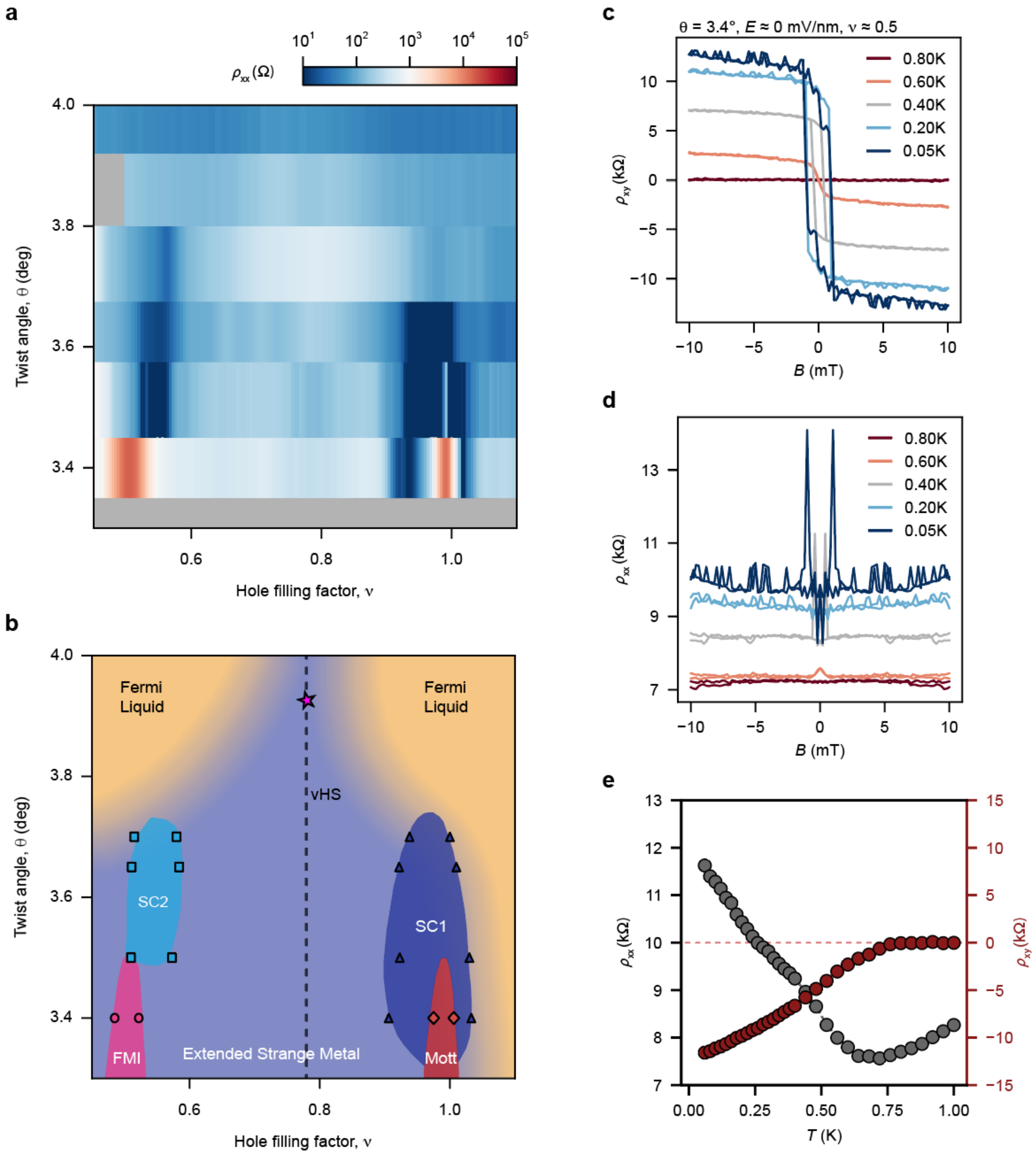

**Figure 4 | Twist-angle-dependent phase diagram at $E \approx 0$mV/nm. a,** $\rho_{xx}$ as a function of $\nu$ and $\theta$ at $T = 30\,$mK and $B = 0\,$T obtained from six different devices. **b,** $\theta - \nu$ phase diagram constructed from the data in **a**. Lavender blue and yellow regions mark the extended strange metal and Fermi liquid, respectively. The black dashed line marks the vHS location; symmetry-breaking states appear below the twist angle marked by the pink star. All other states follow the convention in Fig. **1e**; their phase boundaries are marked by the symbols for each device. **c,d,** Hall ($\rho_{xy}$, **c**) and longitudinal ($\rho_{xx}$, **d**) resistivities as a function of $B$ at selected temperatures for the 3.4° device at $\nu \approx 0.5$. At each temperature, measurements were performed for both field-sweep directions. **e,** Temperature dependence of $\rho_{xx}$ (grey) and $\rho_{xy}$ (red) at $B = 10\,$mT. Red dashed line indicates $\rho_{xy} = 0$kΩ.

## Extended Data Figures

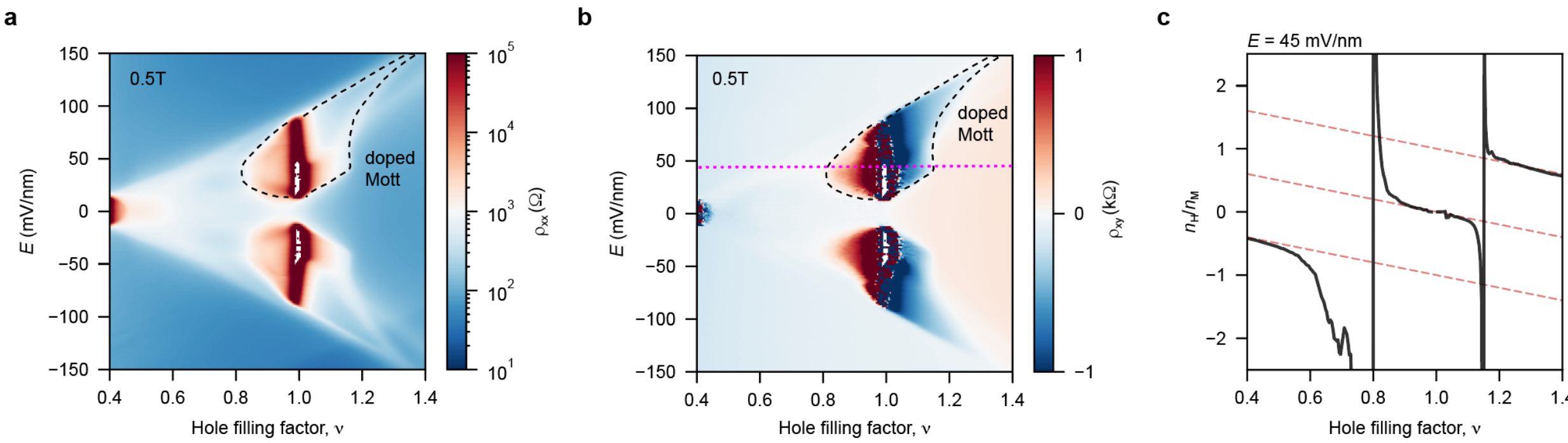

**Extended Data Figure 1 | Hall resistivity map for the 3.65° device. a,b,** $\rho_{xx}$ (**a**) and $\rho_{xy}$ (**b**) as a function of $\nu$ and $E$ at $T = 50$mK and $B = 0.5$T. The black dashed lines enclose the doped Mott insulator region with a reconstructed (small) Fermi surface, where both $\rho_{xx}$ and $\rho_{xy}$ are enhanced and $\rho_{xy}$ changes sign compared to its value outside the enclosed region. **c,** Filling factor dependence of the extracted Hall density ($n_H$, normalized to $n_M$) at the line cut illustrated in **b** (pink dotted line). Dashed lines mark the expected Hall densities at $-\nu$, $1-\nu$ and $2-\nu$. A reset in $n_H$ by about $n_M$ is observed as the system goes from a doped Mott insulator with a small Fermi surface to a band-like Fermi surface ($\nu \lesssim 0.8$ and $\nu \gtrsim 1.2$).

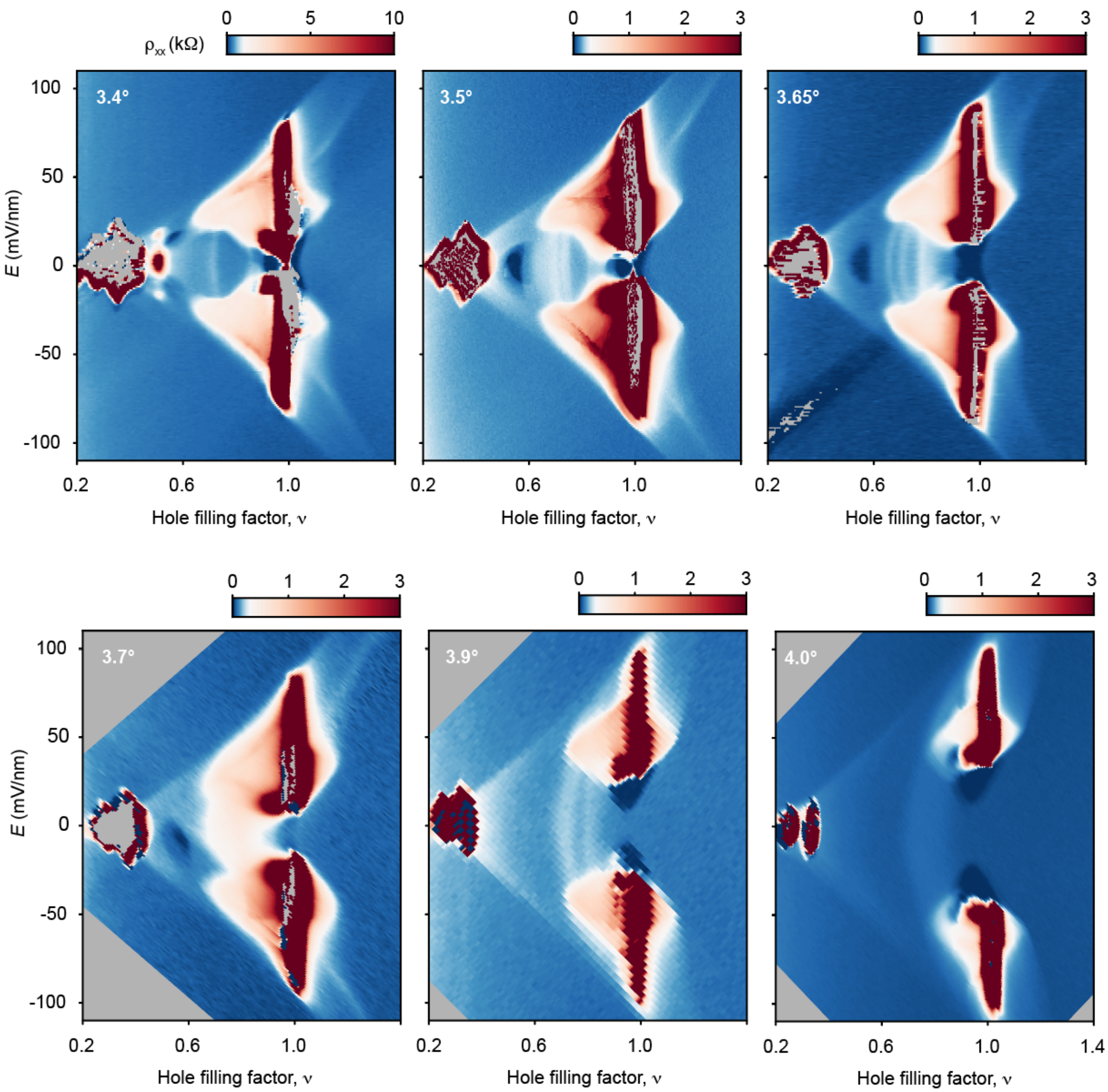

**Extended Data Figure 2 | Summary of twist-angle-dependent resistivity data.** Longitudinal resistivity $\rho_{xx}$ as a function of $\nu$ and $E$ at $T = 30\text{mK}$, $B = 0\text{T}$ and $\theta = 3.4° - 4.0°$. See Fig. **1e** for examples of the constructed phase diagrams.

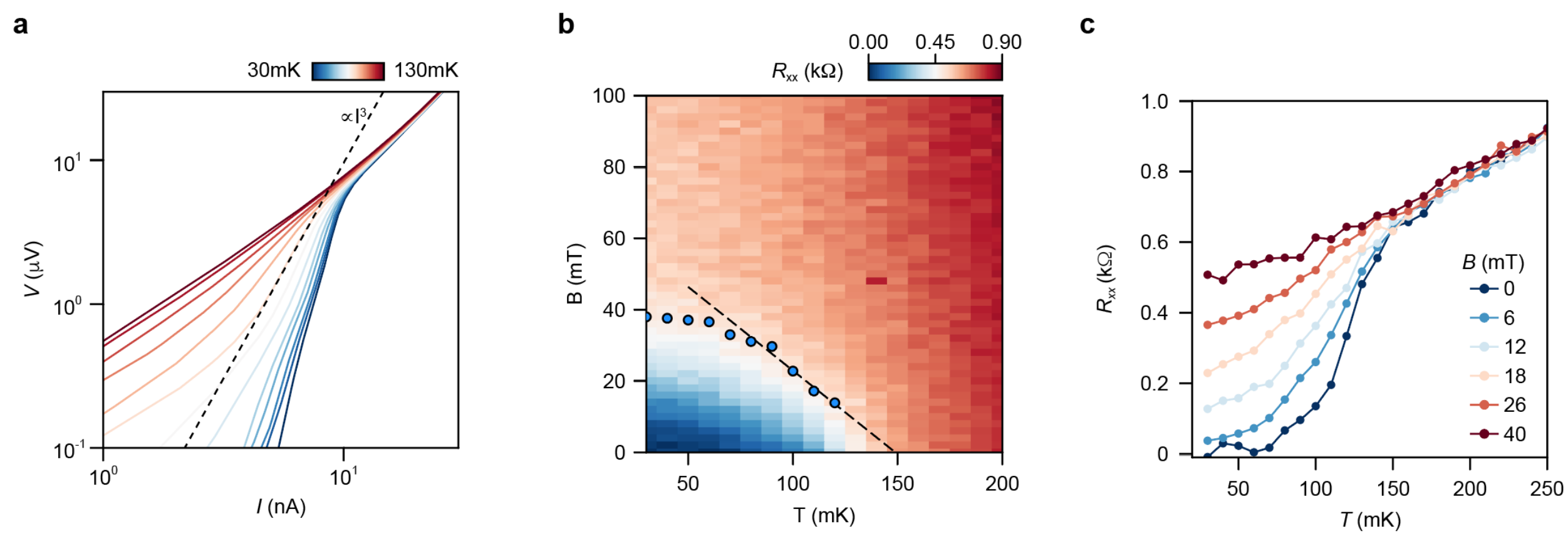

**Extended Data Figure 3 | Additional characterizations on SC2 in the 3.5° device. a,** DC voltage $V$ for the probe electrode pair as a function of the DC bias current $I$ at $B = 0$T. Differently colored curves correspond to temperatures from 30mK to 130mK in 10mK increments. The black dashed line indicates the $V \propto I^3$ dependence, from which the Berezinskii-Kosterlitz-Thouless transition temperature is estimated to be about 100mK. **b,** Zero-bias resistance $R_{xx}$ as a function of $T$ and $B$. Blue dots mark the upper critical field $B_{c2}$ extracted using the Bardeen-Stephen theory. The black dashed line is a fit to the data by $B_{c2} = \frac{\Phi_0}{2\pi\xi^2}(1 - \frac{T}{T_P})$, yielding a coherence length $\xi \approx 69$ nm and a pairing temperature $T_P \approx 150$mK. ($\Phi_0$ is the flux quantum.) **c,** $R_{xx}$ as a function of $T$ at selected magnetic fields. All data were acquired at $\nu \approx 0.55$ and $E = 0$mV/nm.

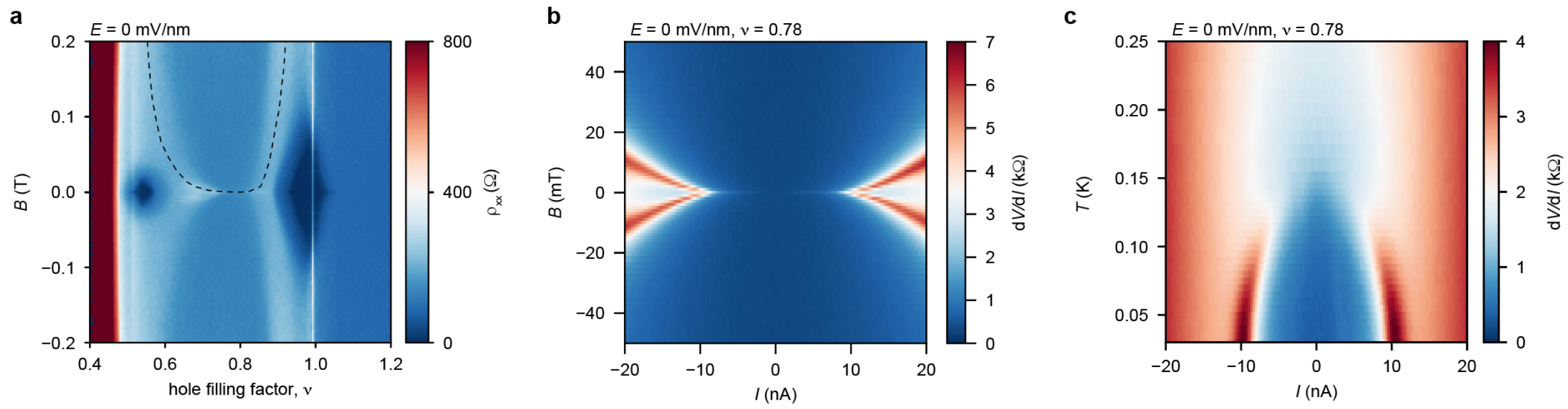

**Extended Data Figure 4 | Symmetry-breaking ground state at the vHS of the $3.5°$ device**. **a,** $\rho_{xx}$ as a function of $\nu$ and $B$ at $E = 0\,\text{mV/nm}$ and $T = 30\,\text{mK}$. The magnetic field quickly quenches both SC1 and SC2; it also quickly splits the vHS (dashed lines) by the Zeeman effect, producing a large negative magnetoresistivity near the vHS. **b,c,** Differential resistance $dV/dI$ as a function of $I$ and $B$ at $T = 30\text{mK}$ (**b**) and as a function of $I$ and $T$ at $B = 0\text{T}$ (**c**). The data in **b** and **c** were acquired at $\nu \approx 0.78$ and $E = 0\text{mV/nm}$. Although the critical current of the symmetry-breaking ground state decreases continuously with increasing temperature (**c**), the increase in critical current with magnetic field in **b** shows that the state is not a superconductor.

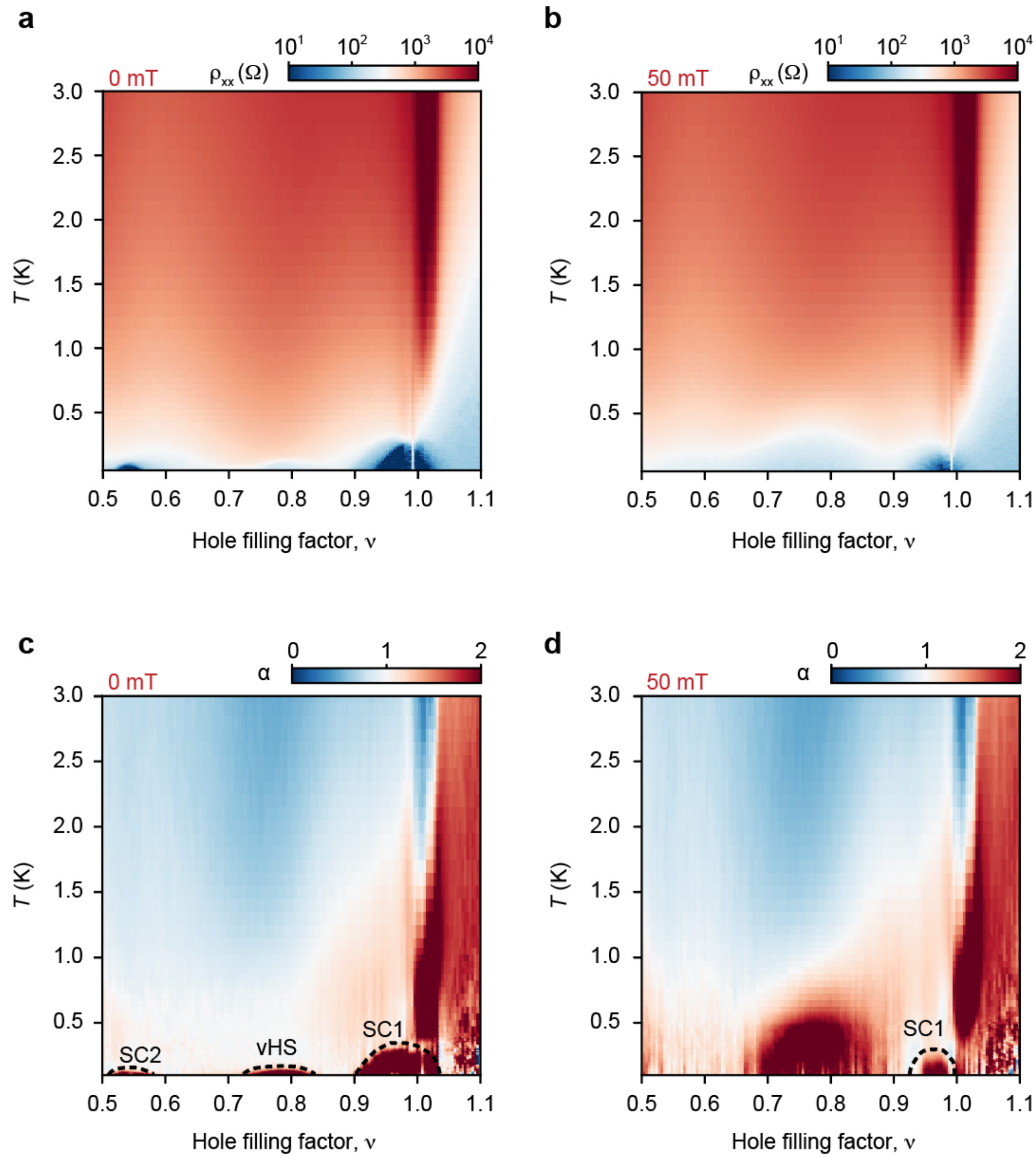

**Extended Data Figure 5 | Extended strange metal under $B = 0$T and 50mT for the 3.5° device.** **a,b,** $\rho_{xx}$ at $E = 0$mV/nm as a function of $\nu$ and $T$ under $B = 0$mT (**a**) and $B = 50$mT (**b**). **c,d,** $\alpha$ as a function of $\nu$ and $T$ extracted from the data in **a** and **b**. The dashed lines mark the symmetry-breaking transition temperatures ($T_c$) for SC1, SC2 and vHS. The data show that a small magnetic field quenches the symmetry-breaking ground state at the vHS and changes the behavior near the vHS [$\nu \approx (0.7 - 0.9)$] from a strange metal (with $\alpha \approx 1$) to a Fermi liquid (with $\alpha \approx 2$) at low temperatures. Note that the extended strange metal survives elsewhere.

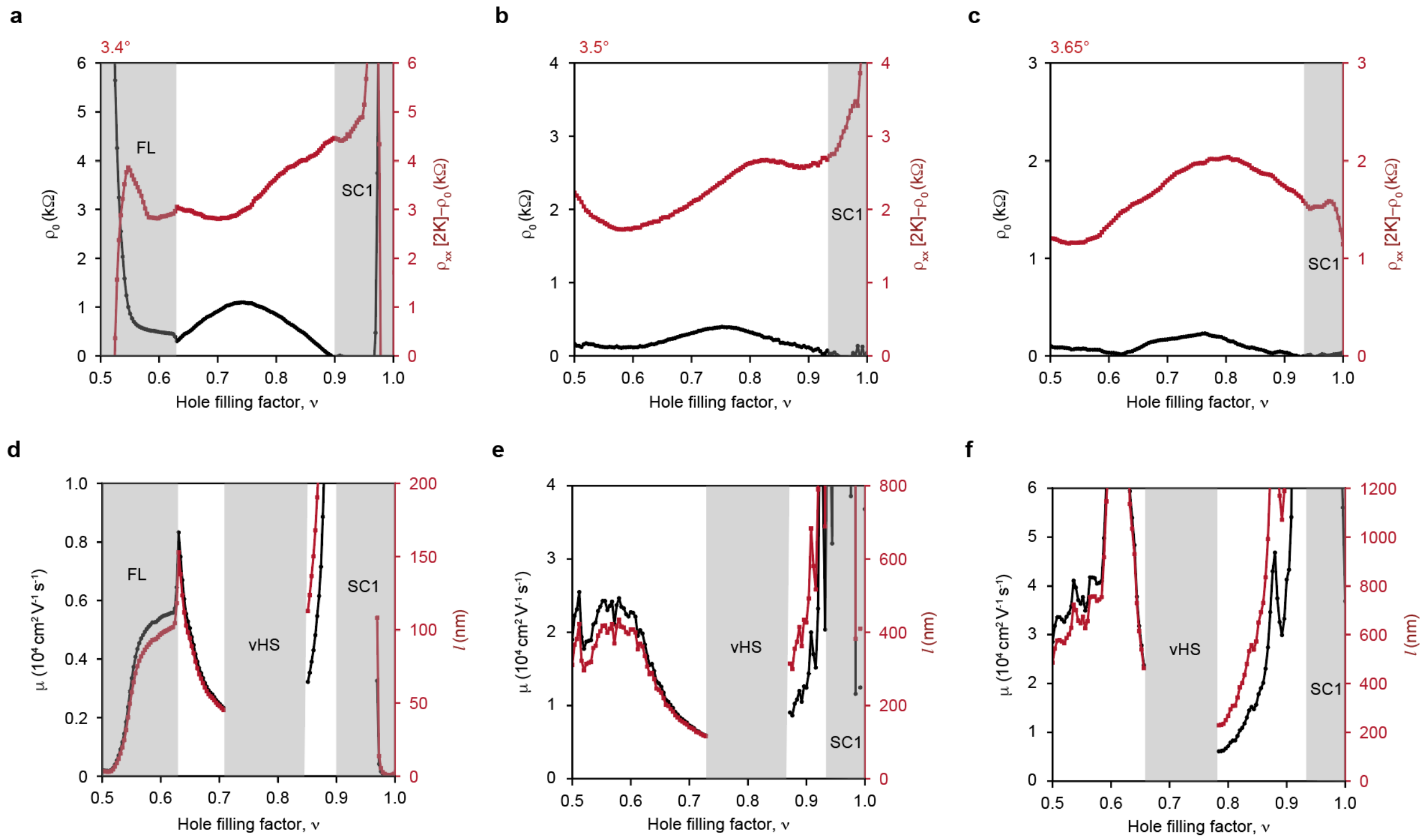

**Extended Data Figure 6 | Estimation of the electron mobility and mean free path from $\rho_0$. a-c,** Residual resistivity $\rho_0$ and resistivity dynamic range $\Delta\rho_{xx} = \rho_{xx}(2K) - \rho_0$ as a function of $\nu$ at $E = 0$mV/nm and $B = 0$T in devices with twist angle $\theta = 3.4°$ (**a**), 3.5° (**b**) and 3.65° (**c**). **d-f,** Electron mobility $\mu$ and mean free path $l$ extracted from $\rho_0$ in **a-c**, respectively. Neither $\mu$ nor $l$ can be extracted reliably near the vHS.

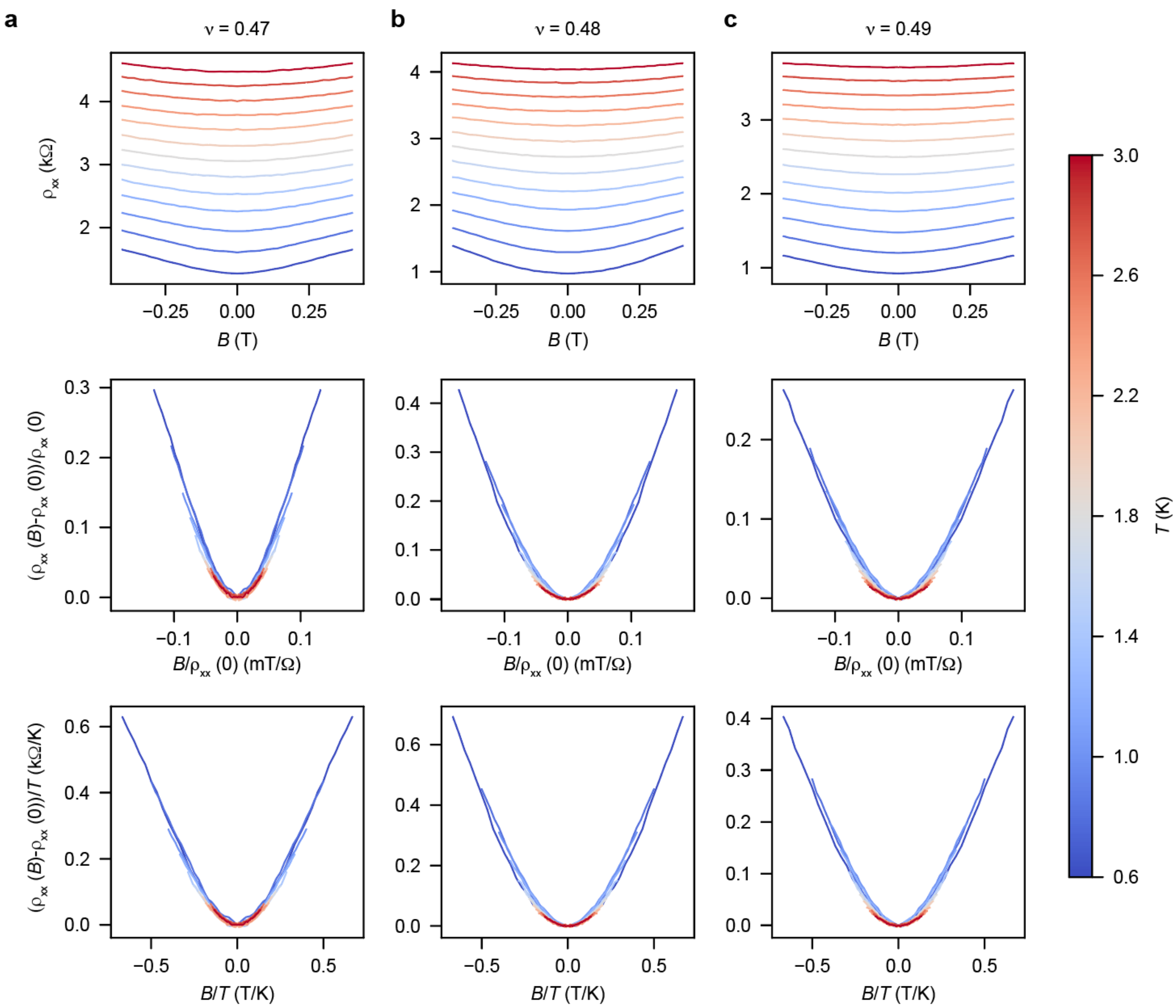

**Extended Data Figure 7 | Kohler and quadrature scaling for magnetoresistivity in the 3.5° device**. **a-c,** Upper panels: $\rho_{xx}$ as a function of $B$ at $\nu = 0.47$ (**a**), 0.48 (**b**) and 0.49 (**c**). Differently colored curves correspond to temperatures from 0.6K to 3K in increments of 0.2K. Middle panels: Magnetoresistivity $[\rho_{xx}(B) - \rho_{xx}(0)]/\rho_{xx}(0)$ plotted as a function of $B/\rho_{xx}(0)$ at the corresponding $\nu$ and $T$, illustrating an approximate obedience to Kohler's rule. Here, $\rho_{xx}(0)$ is the zero-field resistivity. Lower panels: $[\rho_{xx}(B) - \rho_{xx}(0)]/T$ plotted against $B/T$ at the corresponding $\nu$ and $T$, illustrating quadrature scaling. Because of the negligible $\rho_0$, the data appear to follow both scaling rules reasonably well (although quadrature scaling is observed to be slightly better). All measurements were acquired at $E = 0$mV/nm.

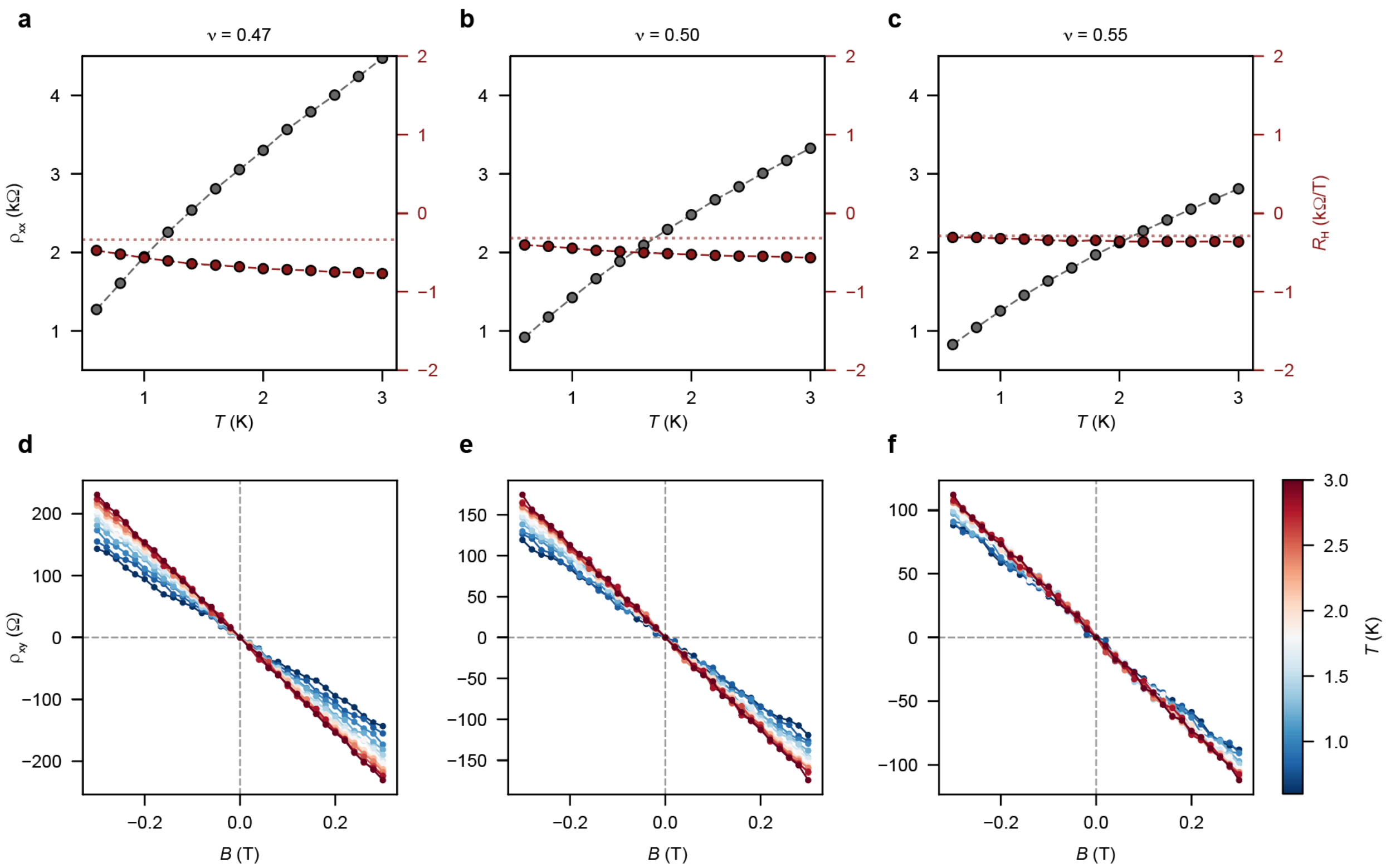

**Extended Data Figure 8 | Temperature dependence of the Hall coefficient for the 3.5° device**. **a-c,** Temperature dependence of $\rho_{xx}$ (grey) and the weak-field Hall coefficient ($R_H$, red) at $\nu = 0.47$ (**a**), 0.50 (**b**) and 0.55 (**c**). Red dashed lines indicate the expected value $R_H = \frac{1}{\nu n_M e}$. $\rho_{xx}$ was measured at $B = 0$T, and $R_H$ was obtained from linear fits to the magnetic field dependence of $\rho_{xy}$ over the range $-0.3\text{T} \leq B \leq 0.3$ T (**d-f**). In **d-f,** curves of different colors correspond to temperatures from 0.6K to 3K in increments of 0.2K. The data show a weak temperature dependence for $\rho_{xy}$, in contrast to the $T^2$-dependence of the inverse Hall angle observed in some strange metals. All measurements were acquired at $E = 0$mV/nm.

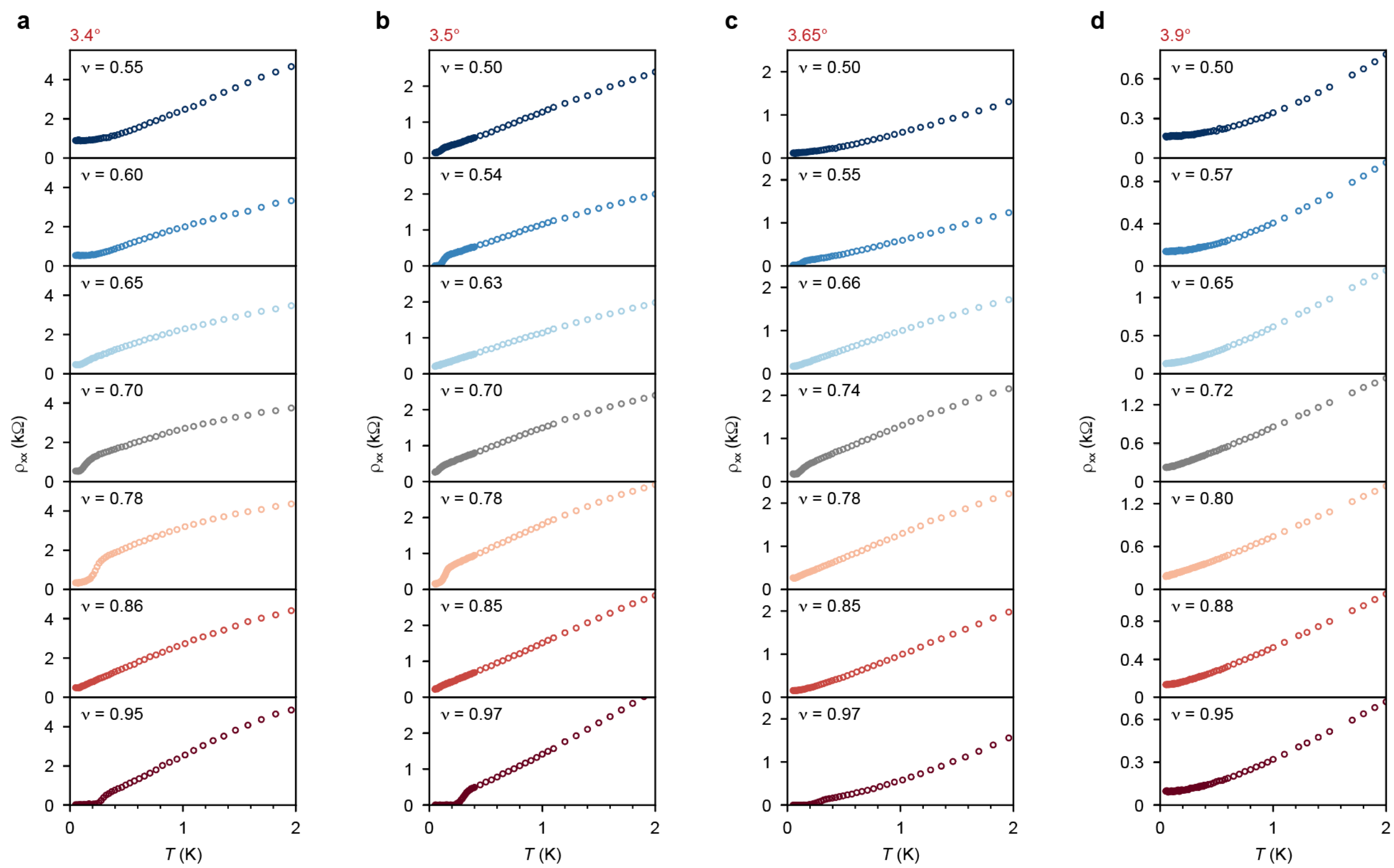

**Extended Data Figure 9 | Line cuts for $\rho_{xx}$ at selected fillings from Fig. 3a-d. a-d,** Temperature dependence of $\rho_{xx}$ at selected fillings under $E = 0\text{mV/nm}$ and $B = 0\text{T}$ for devices with twist angle $\theta = 3.4°$ (**a**), 3.5° (**b**), 3.65° (**c**) and 3.9° (**d**).

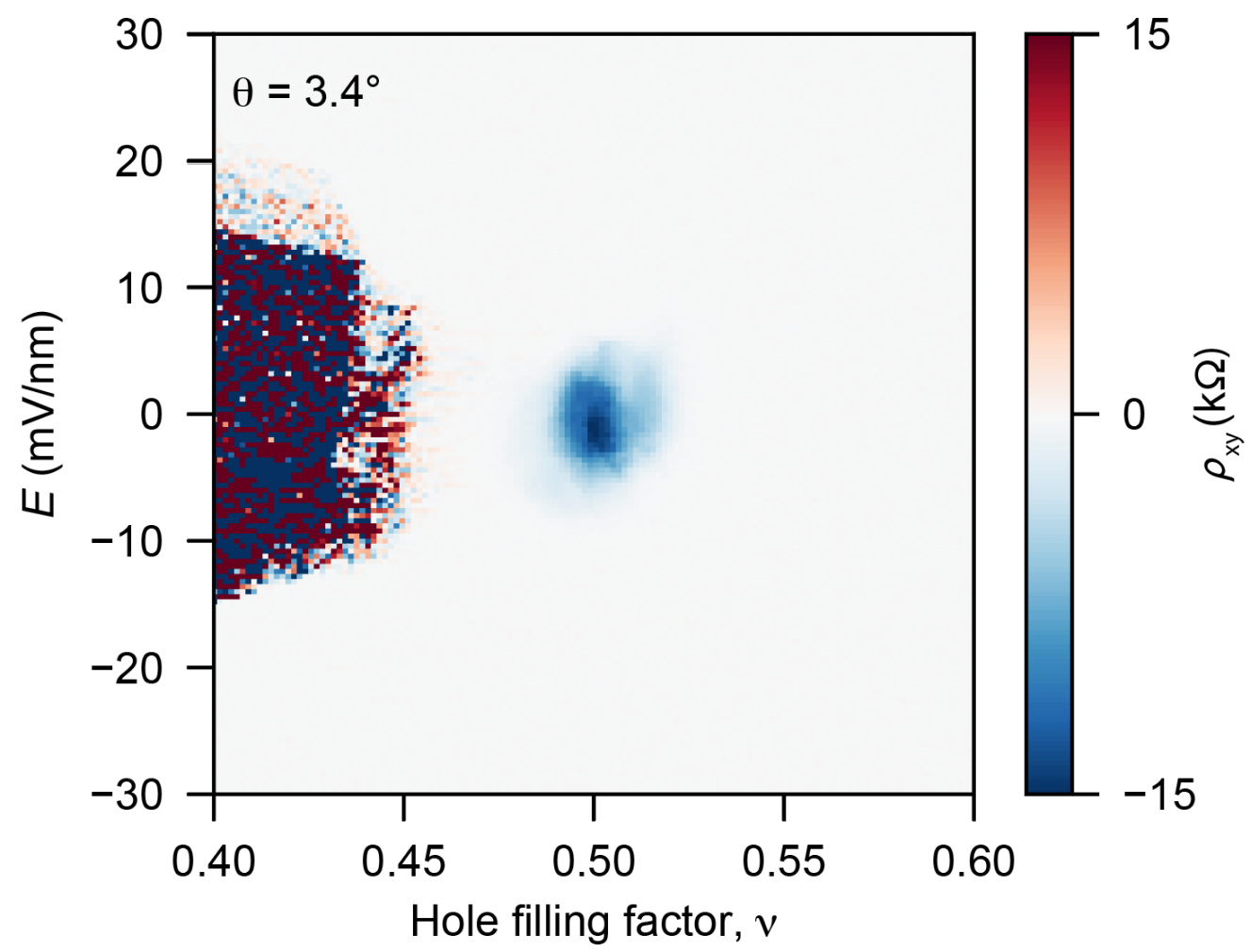

**Extended Data Figure 10 | Anomalous Hall response near $\nu = 1/2$ for the $3.4°$ device**. Anomalous Hall resistivity $\rho_{xy}$ as a function of $\nu$ and $E$ at $T = 30$ mK. The data were antisymmetrized at $B = \pm 10$ mT. An anomalous Hall spot is observed near $\nu = 1/2$ and $E = 0$ mV/nm.

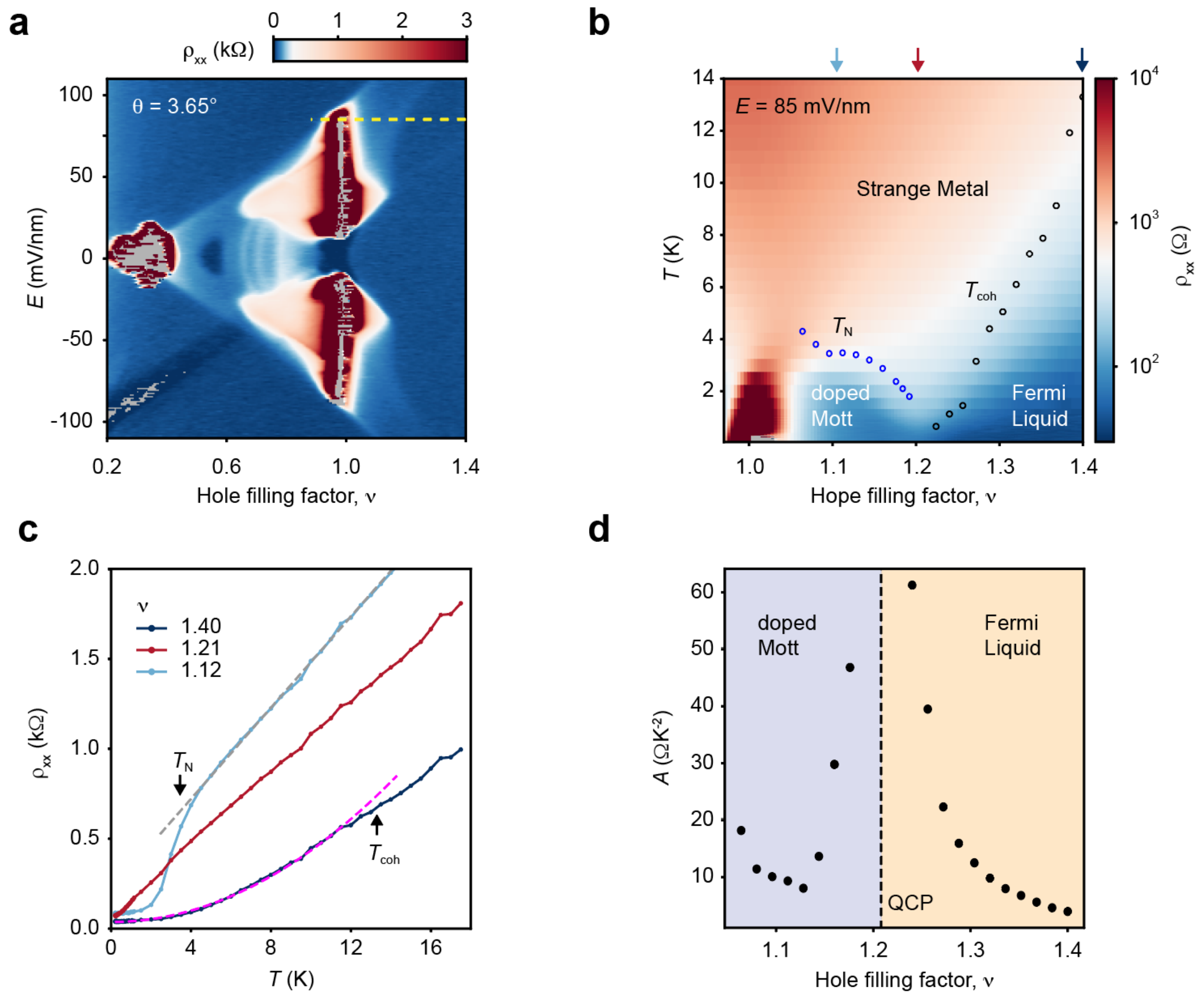

**Extended Data Figure 11 | Sharp quantum critical point (QCP) at a finite electric field in the 3.65° device. a,** $\rho_{xx}$ as a function of $\nu$ and $E$ at $T = 30$mK and $B = 0$T. **b,** $\rho_{xx}$ as a function of $\nu$ and $T$ at $E = 85$mV/nm (the yellow dashed line in **a**) and $B = 0$T. The blue and black data points are, respectively, the extracted transition temperature ($T_N$) for the doped Mott state and coherence temperature ($T_{coh}$) for the Fermi liquid with a band-like Fermi surface. **c,** Line cuts of **b** at selected fillings (arrows in **b**). The grey dashed line is a linear fit to the high-temperature data for the doped Mott state; $T_N$ is determined as the temperature at which the data deviate from the fit by 20%. The pink dashed line is a quadratic fit to the low-temperature data for the Fermi liquid; $T_{coh}$ is determined as the temperature at which the data deviate from the fit by 10%. A strange metal with $T$-linear resistivity down to the lowest temperature is observed at the QCP at $\nu \approx 1.21$. **d,** Doping dependence of the coefficient $A$ extracted by fitting the low-temperature $\rho_{xx}$ data to $\rho_{xx} = \rho_0 + A \times T^2$ on both sides of the QCP. $A$ increases substantially near the QCP. The results here further show that the extended strange metal at $E = 0$mV/nm is not associated with the scenario of a quantum critical region substantially broadened by strong disorder.

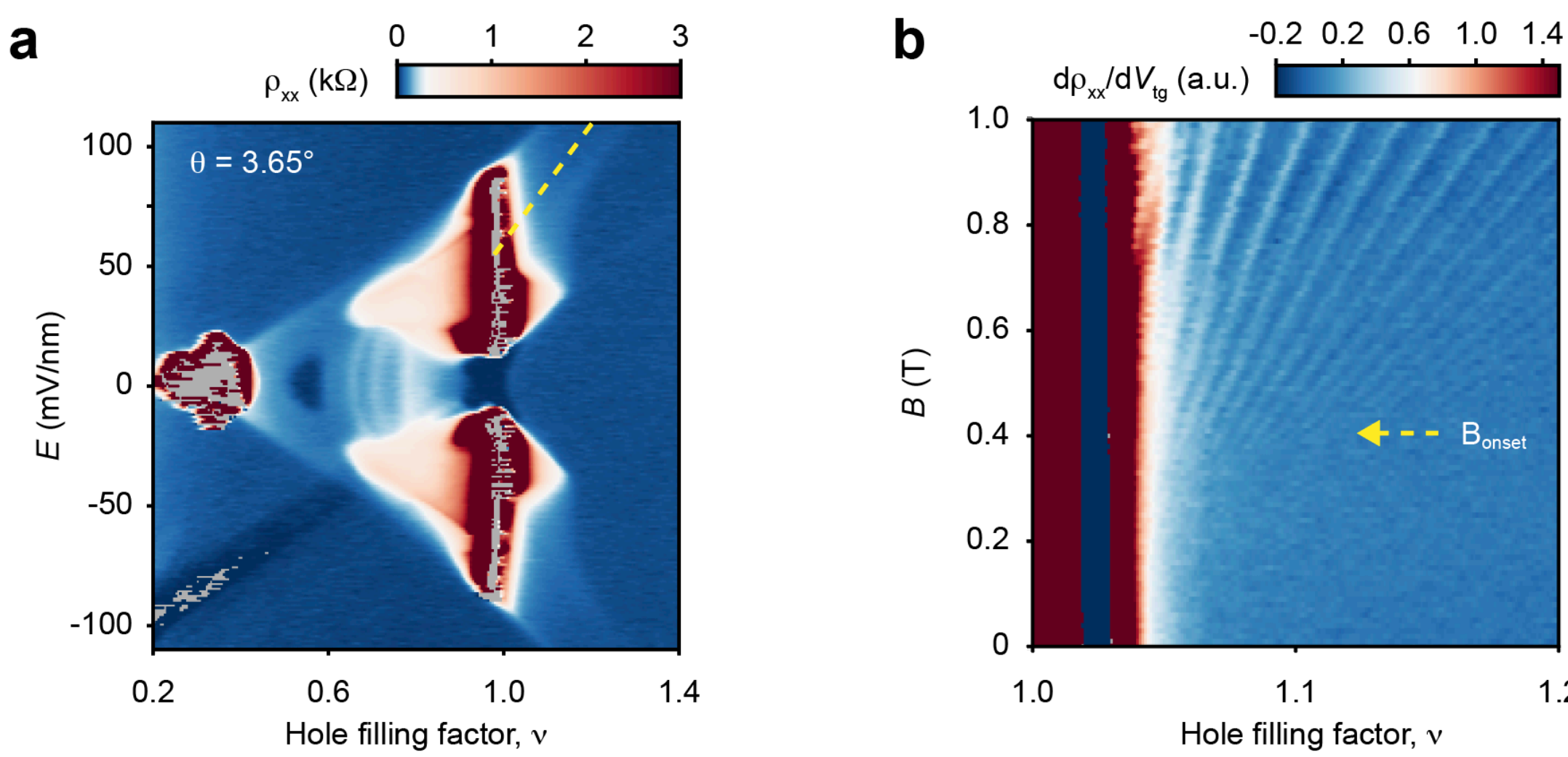

**Extended Data Figure 12 | Estimation of the quantum mobility in the $3.65°$ device. a,** $\rho_{xx}$ as a function of $\nu$ and $E$ at $T = 30\text{mK}$ and $B = 0\text{T}$. **b,** Derivative of $\rho_{xx}$ with respect to the top gate voltage, $d\rho_{xx}/dV_{tg}$, as a function of $\nu$ and $B$, measured along the yellow dashed line in **a**. Shubnikov–de Haas oscillations become discernible above an onset field of $B_{onset} \approx 0.4\,\text{T}$, yielding an estimated quantum mobility of $\mu_q \approx \frac{1}{B_{onset}} \approx 2.5 \times 10^4$ cm$^2$V$^{-1}$s$^{-1}$.